International Conference on New Interfaces for Musical Expression

# Entangling Practice with Artistic and Educational Aims: Interviews on Technology-based Movement-Sound Interactions

**Victor Paredes, Jules Françoise, Frederic Bevilacqua**

**ABSTRACT**

Movement-sound interactive systems are at the interface of different artistic and educational practices. Within this multiplicity of uses, we examine common denominators in terms of learning, appropriation and relationship to technological systems. While these topics have been previously reported at NIME, we wanted to investigate how practitioners, coming from different perspectives, relate to these questions. We conducted interviews with 6 artists who are engaged in movement-sound interactions: 1 performer, 1 performer/composer, 1 composer, 1 teacher/composer, 1 dancer/teacher, 1 dancer. Through a thematic analysis of the transcripts we identified three main themes related to (1) the mediating role of technological tools (2) usability and normativity, and (3) learning and practice. These results provide ground for discussion about the design and study of movement-sound interactive systems.

## Introduction

The study of movement-sound interactive systems, that are designed and developed at the intersection of several disciplines, is one of the important areas of NIME research. First, these systems are conceptualized through embodied interaction and phenomenology, where mind and body are not separated [1]. Second, the development of movement-sound interactive systems is also directly driven by technological advances. Precisely, the wide availability of motion-based technologies has opened new possibilities to either study musical gestures [2], develop Digital Music Instruments (DMIs) [3] or experiment with gestures, movement and bodily awareness in dance [4]. Finally, these movement-sound interactive systems are also driven by artistic endeavors in music, dance, installations, which in turn directly influences how these systems are used and modified.

Interestingly, pioneers such as Michel Waisvisz, Laetitia Sonami, Atau Tanaka, or David Wessel, to name a few, developed and practiced over the years their own instruments/interfaces, even before the NIME conferences started [5]. Since then, many new approaches and interfaces have been proposed and evaluated at NIME conferences [3][5][6][7]. Moreover, as NIME is largely driven by academic research, it often focuses on "new" digital musical instruments rather than on singular practices over time. Understanding how artist build "mastery" of their digital instruments requires analyzing a complex network of questions including skill learning, programming, improvisation and composition, and transmission to others. Several artists who first developed their once novel instruments pursued to develop their own expert practices. While their systems might have been built in collaboration with NIME researchers, the following years of practicing and learning, and sometimes restarting with different systems, might not have always been reported.

For this reason, we wanted to raise these questions and confront them with artists who have practiced with movement-sound interactive systems for several years. The goal of this research is to document through interviews how professionals from different backgrounds (including performers, composers, dancers, and educators) question their practices. Inspired by recent studies based on thematic analysis from semi-directed interviews [8][9], we contacted six different artists to reflect on the following questions:

- which technological systems do they use and why;
- how do they learn, experiment, and practice these systems;
- how do they transmit such systems.

Transmission is meant here as the act of disseminating and/or teaching knowledge and systems to peers or students.

The structure of the paper is as follows. First, we recall some key related works, and we describe the method. The major part of the article consists in the description of three major themes that emerged through the interviews. The results are then discussed with regards to the design of movement-based interactions.

## Related Work

Movement-sound interactive systems rely on different sensor technologies and data processing to map body movement to sound synthesis. Technologies involve motion sensing such as cameras [10][11][12], 3D scanners [13], or Inertial Measurement Units (IMUs) that embed accelerometers, gyroscopes and magnetometers [14][15][16][17][18][19]. Often, instruments do not include visual feedback and users/musicians must rely on their own proprioception. In some cases, haptic feedback (passive or active) can enhance or facilitate the gestural performance. Different levels of control can be designed for the movement-sound interaction: from a deterministic relationship, as found in acoustic-like control, to various degrees of indeterminacy compatible with more exploratory or improvisational control.

By their multimodal nature, movement-sound interactive systems attract various practitioners. Certain performers, more focused on music creation, use instruments in conjunction with IMUs to create augmented play experiences [20] or a combination of motion sensors to create non-tangible interactions with an instrument [21]. In movement-oriented practices such as dance, movers use sound interaction to create installations for people "to discover new interpretation of their dancing body" [22] or foster kinesthetic awareness [23]. Interactive systems can also help music education through embodied interaction of certain concepts such as conductor's gestures [24].

While user studies have been reported on notions such as usability, appropriation [25], or sensorimotor learning [26], we wanted to confront these topics with the view of different practitioners using movement-sound interactions in creation. Reflexive Thematic Analysis, as described by Braun and Clarke [27], is a relevant tool for theme extraction from qualitative data such as interview transcripts. It has been used to extract mapping design criteria in NIME [8], musical use habits of mobile devices [28] or for interviews with sound artists around a given practice [9]. Yet, studies focusing on learning and long-term practice remain scarce within the NIME community.

## Method

Following the approach of Robson et al. [9] and West et al. [8], we conducted semi-directed interviews and thematic analysis to better understand how experts relate to questions of learning, practice, and relation to technology.

### Participants

We contacted 6 artists that used and/or designed movement-sound interactive systems in their practice. All artists collaborated or were in contact with the last two authors, but not the first author who led the interviews. All artists accepted to participate in the study, and give the consent to the publication of the article. We led three interviews at the beginning of 2021 and the rest at the end of the year. We ensured gender balance and covered various practices: we interviewed 2 music performers (Mari Kimura and Philippe Spiesser), 2 dancers (Bertha Bermudez and Yves Candau), 1 teacher (Fabrice Guédy) and 1 composer (Michelle Agnes Magalhaes). A description of the interviewees' background is available in appendix.

### Study Design

This study aimed to investigate how practitioners engage with learning and appropriating technologies that mediate movement and sound in artistic or pedagogical contexts. The interviews were semi-structured around five topics with a set of questions that could be adapted to the practice of the interviewee: (1) background and motivation in using movement sensors; (2) the process of learning movement and/or sound; (3) the transmission of such systems; (4) technology appropriation (by performers and public); (5) limitations and perspectives. A list of the generic questions asked is available in appendix.

Two of the interviews were led in person and the rest over video conference. Interviews lasted between 90 and 120 minutes. Interviews were automatically transcribed first using Microsoft Speech to Text service, and then manually corrected using the audio recording.

## Analysis

We decided to do a reflexive Thematic Analysis (TA) on the transcripts, as described by Braun and Clarke [27]. TA is a qualitative data analysis method, well established in the social sciences and psychology, that enables researchers to extract patterns of meaning called *themes* across a dataset. It consists, after familiarization with the data, of a first phase of coding and a second phase of theme development. Reflexive TA differs from other TA techniques in that it acknowledges the subjectivity that lies in qualitative data interpretation. The theme development emerges from the interpretative work of the researcher made *from* the codes. The codes do not come from a codebook but are *generated* by the researcher when analyzing the data. The themes should not be defined prior to the coding and must emerge as patterns from those codes. Moreover, reflexive TA is "theoretically flexible", not bound to certain methodologies that were initially meant for psychotherapy research and psychology. It applies to a broad range of domains and more generally suits the exploration of participants' subjective experiences [29].

We all individually coded the transcripts and regrouped them under potential themes. We then converged together toward three main themes: (1) Practitioners consider technology as mediating practice, (2) they navigate usability and normativity, and (3) their learning process is entangled with the development of a personal practice.

## Theme 1: Technology Mediates Practice

During the interview we explicitly asked the artists to position their work and approaches with respect to the use of *technologies*. We did not provide any definition of what technologies could refer to, and let them describe how they situate themselves with this large topic. We can describe their discourses by *what* technology is, *why* using it and *how.*

## What is Technology, Why Use it and How?

First, they either use the term "technology" in a very general sense (which could be induced by the fact we use it in questions), or they refer to the specific tools they have been using in their work (such as the Kinect, smartphones, etc). They do not refer to specific categories as used in engineering fields such as motion capture, machine learning or AI, to cite examples of categories currently widely used. Interestingly, YC even considers *"A somewhat extended notion of technology. As one could say that language is a technology, that writing is a technology"* (YC).

Importantly, technology is neither considered as a "theme" or "goal" in their practice, but rather as something embedded in a socio-cultural context, e.g. as *"[musical] instruments with today's technology"* (PS). They all consider that integrating technology in a personal practice is a complex process that demands significant engagement to become a meaningful part of the practice as opposed to development made *"for the sake of technology"* (BB). By considering technologies an integral part of the practice, thus as a mean of expression, such an approach contrasts with other fields involving research and development which might prioritize novel technologies within a solutionist perspective.

Technology is therefore never thought of as a "solution" to a given problem, nor as a mean to realize an artistic vision. It is rather seen as a beneficial "perturbation", enabling them to *"move out of their comfort zone"* (PS), or something they have to deal with. In all cases, it should be grounded in a methodological approach, that might need to be invented. Possibility of looking for systems failures or finding inspiration in bugs are also explicitly described.

> I like when there are bugs, because it's in these interstices that there are often opportunities to find things that you wouldn't have found if everything had been done beforehand (FG).

## Technology Can Act as a Mediator for Transmitting Concepts and Ideas

Most artists also considered interactive technologies as a support for transmission. The most prominent example is FG who uses, among other things, movement technologies to help teach music theory to young students. Using those technologies is a way to create bridges between tradition in music education and the world of computer science, and *"it is very clear that the de-compartmentalization that it produces in students is something that works."* (FG). BB has never experienced movement interaction technologies as *"a tool for the process of creation to be enhanced. But for me, it has been a tool really for transmission."* A tool that she uses to initiate and teach non dancers movement practices. For BB, working with interactive systems requires producing a synthetic model of the practice, which benefits its transmission:

> It helped me to understand […] how to make it accessible. Because reducing, summarizing, also helped us change the transmission, develop a clearer terminology, be more conscious, really, of what we make. (BB)

### The Mediating Role of Technology Might be Unclear for the Audience

Performances with movement technologies are still new to the general audience. YC, PS and FG reported that there was an effort to be made towards the understanding and/or acceptance of this interaction by part of the public. PS and YC pointed out that

> There is this attitude, as a first response to a new system, of wanting to understand what is going on. (YC)

The technology becomes central in the performance, overshadowing the artistic proposition. For MK, however: *"if you have the motion sensor that is moving, people can actually see and relate to the electronic sounds better."* In this perspective, the interaction actually links the audience to the produced sound. But further work is needed for spectators to adapt to the interaction, particularly with non tangible interfaces, where the underlying link between movement and sound is blurred :

> That is, the smallest gesture you make becomes an identifiable and interpretable musical gesture for an audience. (PS)

### Theme 2: Navigating Usability, Normativity and Obsolescence

As interviews revolved around technology, interviewees offered insights regarding the usability of interactive systems. In particular, the potential and ease of use of some technologies and feedback modalities over others was discussed. Critically, the question of the normative power of technologies over their users was a recurring theme of the interviews, as well as issues of obsolescence.

## Technology Requires Commitment and Friction

Most participants have a history of practice that involves the use of wearable sensors embedding IMUs. They are appreciated because they are easy to set up, are minimally invasive and have a low latency. In particular, mobile phones greatly facilitate setup. While usability is appreciated, interviewees still emphasized that working with technology takes time and requires dedication, engagement and commitment. Reflecting on a decade of experience, MK reported:

> I've practiced well enough using just one thing. […] But in order to refine your movement and try to incorporate that into a tool that's useful for musical expression, you really need to spend time with it. (MK)

Issues such as calibration and parameterization have a learning curve but are critical for both performance and transmission. All interviewees emphasized a desire to escape a triggering paradigm found limited and often associated with a "gadget". Yet, apprehending more complex forms of interaction can be difficult. For PS, as complexity increases, *"you can get overwhelmed very quickly because you become a human synthesizer and so all you do is sound."* Apprehending such increasing complexity is not easy and interpreters need to learn to play with the subtleties of movement. Yet, interviewees often valued the friction that occurs over the development of a practice:

> Because the tool resists, the user will persist a little bit in doing it and that's where he'll find things that I find interesting. Typically, in the gesture tracking, […] it wouldn't have worked if it had worked right away. It's a paradoxical way of putting it, but I think that's what can make you think bigger. *(*FG*)*

## Normative Technologies should be Avoided

Practitioners value friction — to some extent — as a mechanism going against potentially normative effects of technology. In creative practices, technologies should not be designed to solve problems and facilitate use to the extent of *"infantilization"*. What motivates FG is *"this interweaving":*

> We are not going to use ready-made technology. […] What is interesting is to associate it from the start, to think about several disciplines at the same time, which will then be mixed together. You're not going to be interested in transmitting a technological brick, but really more the process of reflection behind the creation. (FG)

Several interviewees reflected on the history of technology in art, drawing comparisons between early experiments with technologies that aimed to broaden the scope of possibilities, and some contemporary technologies that have become normative. Most interviewees make efforts to foster creativity and personality in their transmission of the technologies rather than a form of imitation and replication:

> We do our best to reproduce the best performance but as soon as we think like that, everything is screwed up because we will put ourselves in a mechanical mode. […] And as soon as we are there, focused on the result, we are no longer in the present. (MAM)

In creation and pedagogy, technology therefore aims to emancipate students and support the development of a personal practice rather than help them reproduce an "ideal" result.

## Obsolescence

The notion of obsolescence consistently emerged throughout discussions with all practitioners, even though it was not introduced by the researchers as part of the interview structure. With the rapid pace of technological innovation, interactive systems can rapidly become outdated and unusable. Such obsolescence creates tensions among artists with regards to the preservation of creations.

> Ligeti didn't want to go into interactive because he thought that the technology will be obsolete, and he's correct, that if you write in a certain way and if that technology dies, your piece dies. (MK)

For MK, such obsolescence is not problematic because her works, her creations will survive. Yet, obsolescence can make transmission hard, as platform updates might prevent students from using the software. For BB, what matters is the persistence of methods and knowledge over time rather than the technology itself:

> The installation, maybe if we put it back it doesn't work anymore. It's not a maybe, it is the reality. […] That's a problem, you could say, it's really a problem because the tool itself is obsolete. But not the creation process. What did we do during the creation? That's not obsolete, because we're still here. So we have to talk about that, we have to leave a trace because I think it's important, I think it's quite unique. (BB)

## Theme 3: Learning and the Development of a Personal Practice

Learning was one of the key themes introduced in the interviews. Our analysis revealed key learning processes and insights into how technologies can be designed to structure movement learning. Importantly, interviewees emphasized that learning is highly intertwined with the development of a personal practice.

### Discovering, Experimenting and Challenging Limits

Approaching a new form of interaction involves several phases of discovery, exploration and integration. When experimenting with a new interactive system, people tend to rely on their existing skills:

> So I think in a learning system, you always start with the things you know first and try to apply them to a new environment. I tried to look for those gestures that I knew. For example, in this virtual instrument system, I imagined hitting a skin, I imagined hitting a block, shaking the maracas, doing a deathstroke. And so I tried to reproduce these gestures, I started from there. (PS)

In this early stage, the quality of attention is essential: the mover must remain open to the feedback provided by the system to get acquainted with a new action-perception loop. This involves moving while listening to the system's responses, as emphasized by MK: *"You know hand movements or body movements, but then use their movement to incorporate the sensors and the first thing I usually do with the students is that I have them wear it right now [...] and see if there's any relevant thing that you can use as an expression"*. Often, as previously mentioned, this exploration is initially driven by the desire to understand the interaction at play.

Naturally, all interviewees emphasized the need for experimentation and exploration. FG, describing learning in a pedagogical context: *"At the workshop, we try without having a clear goal, but we know that it is still interesting to try. [...] In fact, the desire itself often refers to something that we want to see, like a child who, even if he doesn't know what the toy is for, will try to throw it"*. Whether practicing for a choreographed or improvised outcome, many participants mentioned improvisation as a key marker in the learning process. Improvisation can be seen both as a mean to explore the capabilities of an artifact and a form of challenge to oneself, that assesses their ability to interact and express with the system. Such experimentation involves exploration and exploitation, and repetition is key in acquiring embodied skills: *"Through the repetition of the same gesture, you can learn, and that's exactly what we do. So you can learn from repetition, that's what you do in a dance studio."* (BB). Experimentation is considered essential to explore the instrument but also as a process that extends beyond the use of the technology itself:

> It's the very activity of coding that is interesting, to think in a certain way. Afterwards, we go back to the piano, to the violin, but we have understood things differently, and we have explored a whole space. That's what it allows you to do, to experiment and to try, but not with the aim of necessarily making an application." (FG)

Once the initial phase of discovery leads to an understanding of the interaction, performers often challenge the system, exploring its limits to understand where it breaks.

> One interesting thing is that we always try to do two things: one, we understand the system, so we do it, we are obedient, […] and then we try to break the system, and see when it doesn't work. […] Almost all the dancers did the same. (BB)

Finding limits involves exploring the potential and shortcomings of the technology, but it is also seen as a process of one's exploration of their own abilities and habits. Technology *"helps me to understand the material better and to get out of a habitus that normally I would have in the studio"* (BB).

## Technologies can be Designed to Facilitate Learning

The way interactive systems are designed can affect their learnability, in the context of open-ended learning processes that integrates with a personal practice. Several factors that could inform the design of learnable interactions emerged through the interviews. First, as highlighted above, constraints provide valuable stimulation for learning, for they drive attentional processes and challenge the performer's habits. Systems can be designed to stimulate movement through metaphors and images, or more implicitly so that movement emerges through interaction.

> The identification of sound-movement […] will be learned by the body, but I don't need to verbalize it. He will understand if he turns it like this that it won't work because it's not the right sound. But I don't need to write a detailed instruction for that because the own object, the device itself will indicate that it's not in the right direction. It's also about trusting the body. The body is not stupid. (MAM)

For PS, interactive systems can be designed for learning by relying on play: *"[a child] will learn gestures very quickly, by himself. So if we can frame it, find an interaction,* *it's a bit like video games. You go through different stages."* However, several interviewees mentioned the need for adaptability, so that *"you make your own path"* (PS). It is important to find a balance between *"a technology that adapts to the interpreter"*, and the need for interpreters to adapt to, learn from and appropriate the technology.

> It's not a tool that I could pass on to anyone. It's a technology, it's a way of seeing things. After that, it's up to you to adapt it, to imagine an interesting use. (FG)

## Learning is Intertwined with the Development of Personal Practice

Reflecting on the details of one's learning process seems to be challenging, because contrary to other fields involving movement learning such as sports, that are task-oriented, expression is often the primary purpose. Learning appears to be inseparable from the development of a personal practice: quoting BB, *"I think that at the level of movement acquisition or understanding it was not about the form or the execution, but rather about the personal practice, how you work."* Depending on the focus of the interviewee, technology can mediate with the theoretical underpinnings of the practice, can foster movement practice and kinesthetic awareness, or can support creativity. When technology is used to support an embodied practice such as dance, its reflective qualities can help mediate embodied experiences, helping the mover become aware of their body, of particular movements, of felt experiences:

> It helps to be conscious of what you do [...] because you have a response, the movement becomes almost tactile. It becomes tangible. The movement has a trace, either in sound or in images, that reflects what you just did. (BB)

As such, the learning process is not merely about gaining pure control over the instrument, but about developing skills within a personal practice, by drawing upon the shifts in the lived experience induced by an external feedback. Sound feedback brings attention to the movement from another modality, therefore structuring experience.

> Technology can structure learning but can also structure experience. There's something a little more marked, with fewer dimensions, and that allows you to channel the experience. (YC)

Technology can therefore shape movement but it affects perception more broadly, quoting MAM: *"what I really want is for musicians to come out of an experience like that transformed. That they can act on their field of perception"*. As a result, most interviewees view their interactions with technology beyond the pure control of an instrument, as a reflective and conversational process. Technology is considered beyond a tool or an instrument, as *"a partner, who helps me to think about what I am doing"* (BB).

What characterizes the entanglement of learning and practice development is the search for a compromise between freedom and constraints. The use of extrinsic feedback on movement brings constraints that are found stimulating for learning, as a way to structure attention, because *"There is a big difference when it's a personal choice and when it's a choice that comes from outside. And also the personal choices, often, happen more automatically. So there's really a gain in having choices that are made from the outside"* (YC). For YC, changes and ruptures of the interaction over time were found particularly stimulating:

> I would say that the things that are very strong in the experience are moments of surprise, for example where there is something established that I can interact with, and I develop some sense of what can be done, and then all of a sudden there is something else that manifests itself, that appears in response to something that I am doing, but that was not there before. (YC)

Such external constraints drive performers out of common patterns and habits. Yet, technology is used by artists to support an expressive practice, and all interviewees expressed the need for a certain degree of freedom: "*it was important to find myself as an interpreter. I trigger things, but I also have the possibility to be free between some triggers, to feel breaths*" (PS). Such freedom is necessary so that interaction lets personality and idiosyncrasies emerge among different interpreters: *"Obviously the piece is totally different with the same parameters. That's what they can influence, they can put their personality without having to change the system exactly"* (PS).

## Discussion

### Expressivity in Question

An important question raised by the NIME community, at least at its start, concerns the notion of "expressivity" — which is included in the NIME acronym [30][31][32]. The use of embodied interaction involving movements and the body for music performance certainly questions how musical expression can be conveyed to the public. Interestingly, even when explicitly questioned, the notion of expressivity was regarded as somehow ill-defined or not considered as an important topic to discuss frontally. Therefore, it would seem that the use of gestures is not meant to enhance or facilitate expressivity, as one could naively believe.

The notion of "expressivity" appeared indirectly, when the interviewees, especially the ones concerned with music performance, did mention spontaneously that the public reception could be problematic. The "expressive intention" behind the movements and gestures might not be always perceived by the audience, especially if they are unfamiliar with such approaches of technology. "How should spectators experience a performer's interaction with a computer?" [33], Reeves and al. questioned nearly two decades ago. The reactions of part of the audience raise a more general question: who are sound-movement interaction systems designed for ? It seems that they grow from the need of pushing certain boundaries in one's practice, and therefore help interact differently with sound or gesture, the finality remaining the same: nurturing creativity. So when transitioning from the studio to the stage, the technology should not be the center of attention. The use of new technologies on stage requires thinking about the perception of the performance by the audience.

To face this challenge it is possible to use spectator experience augmentation techniques (SEAT), that helps improve the audience experience, as shown by Capra et al. [34]. However, "the word 'audience' suggests a passive role, whereby you are rewarded with culture simply by virtue of turning up" [9]. Robson et al. have interviewed sound artists who are engaged in situated sonic practices. One of the interviewees noted that "[Visitors] need to do something to bring the whole installation together, move around and fetch different parts."[1] . Similarly to sound installations where receivers must actively make sense of the piece by changing their spatial relationship to it, it would be beneficial in the process of "restoring trust with the spectators" [34] to help get them from being part of a passive audience to active receivers, not of a technological demonstration, but of an artistic proposition.

## Questioning the role of technology

In all interviews, the view of technology is ambivalent, navigating between normativity and emancipation. While it is considered central to current artistic practices, as part of a global socioeconomic system, it is also seen as a "tool" to experiment and create beyond norms. This creates an interesting paradox, since the need for experimentation might go against the establishment of standards that are necessary for community building and transmission. Moreover, obsolescence is unavoidable, based on evidence of using technologies for several years, and yet the general concepts that are developed are intended to resist technological evolution. This brings important questions about transmission and pedagogy, some of which have been brought up by several researchers [35][36][37]. We believe that it opens a large and interesting debate, since this necessitates to establish what in our practices is truly independent from technological artifacts.

## Supporting Learning

It appears that pedagogy inherited from standard music practices does not encompass the variety of musical practices that exists today. Particularly in improvisation practices, as shown by Hayes [38], the notion of graduating from novices to expert with a predefined educational path does not apply, as it puts the skillful musical instrument expertise before the value of "the instantiation of multiple sensitivities of the person as a whole" [38]. The role of interactive systems is fundamentally perceived as elements that should be sufficiently modular, and assembled in different manners. As such, interviews highlighted that systems can be designed to facilitate learning, through metaphors, images, play, and interactive feedback supporting movement execution. This can be linked to the use of different types of affordances as suggested by Altavilla et al. [39]. Because such systems alter the action-perception loop, they bring external constraints that structure the lived movement experienced, and stimulate attention and kinesthetic awareness. Yet, designing systems in a way that eases the learning process involves a number of challenges. Crafting the right level of detail for movement-sound interactions can hardly be optimized for all learners, because the learning process is open-ended and intertwined with the user's own personal practice. As such, interviews show that performers often co-evolve with the system, progressively uncovering subtleties in movement expression. Designing such adaptive systems seems promising but requires delineating what should change and what should remain consistent within a learning process that is always personal.

## Conclusion

The interviews we conducted were found fruitful to bring a different light to NIME research about movement-sound interactive systems. As discussed by other researcher in this field, it is intrinsically difficult, or maybe fundamentally impossible, to translate global methods borrowed from traditional music practices, in term of composition and pedagogy, to practices based on interactive technology. Nevertheless, the insights given by the artists still reveals how important it is to privilege strong concepts over detailed implementation, the importance of openness and modularity of the system, and how constraints and perturbations can be fruitful in artistic contexts.

## Acknowledgements

This research was supported by the ELEMENT project (ANR-18-CE33-0002) from the French National Research Agency. We warmly thank all six interviewees for the time they allotted us and the openness with which they answered our broad questions.

## Ethics Statements

All participants were contacted by mail, prior to the interviews, explaining the aim and the course of the study. Participants all gave their informed consent to participate to the study and to be recorded for later transcription and analysis, and agreed to appear in the article non-anonymously. Moreover, we ensured gender balance of the interviewees.

## Appendix

### Participant's background

We introduce here the artists who participated to the study along with a short description of their practices and associated works that were specifically referenced in the paper.

**Bertha Bermudez**[2] (BB):
*Dancer, Dance Educator*
*Selection of works:* Double Skin Double Mind Installation [22][40], CoMo-Elements [41]

**Yves Candau**[3] (YC):
*Dancer/artist, Coder*
*Selection of works:* CO/DA, Still Moving [23]

**Fabrice Guédy**[4] (FG):
*Music Educator, Composer, Pianist*
*Selection of works:* Gesture Follower [24], Modular Musical Object [42], Concert Féminin / Féminine[5], Volière[6]

**Mari Kimura**[7] (MK):
*Violinist, Composer, Music Educator*

*Selection of works:* Augmented violin with sensors [20], MUGIC[8]

---

**Michelle Agnes Magalhaes**[9] (MAM):
*Composer, Performer, Music Educator*
*Selection of works:* Constella(c)tions[10], CoMo-Elements [41]

---

**Philippe Spiesser**[11] (PS):
*Percussionist/Performer, Music Educator*
Selection of works: GeKiPe [21], SkinAct[12]

## Questions

The questions used during interviews vary according to the interviewee's practice but the formulation and general themes remained consistent throughout the study. The following set of questions has been used to structure the interview with MK :

1. How would you describe yourself and your practices today?
2. What motivates you today in using technologies for movement in your artistic practice?
3. How do you create, practice and learn new pieces involving technology?
4. Would you like to pass on the knowledge and experience you have on these technologies to others? How?
5. Do you feel that your personality as a musician comes through when you play with this system?
6. What is your point of view about expressivity using technology / motion sensing? Has it evolved over time?

## Footnotes

1. Quote from the interview of Roswitha von den Driesch and Jens-Uwe Dyffort, Sound artists from Berlin, Germany. ↩
2. https://www.lafaktoria.org/en/bertha-bermudez/, accessed 15th April 2022. ↩
3. Vimeo page, accessed 15th April 2022. ↩
4. https://feuillantines.com/, accessed 15th April 2022. ↩
5. https://www.bnf.fr/fr/agenda/concert-feminin-feminine, accessed 15th April 2022. ↩

6.  [http://gallicastudio.bnf.fr/voliere](http://gallicastudio.bnf.fr/voliere), accessed 15th April 2022. ↩

7.  [http://www.marikimura.com/](http://www.marikimura.com/), accessed 15th April 2022. ↩

8.  [https://mugicmotion.com/](https://mugicmotion.com/), accessed 15th April 2022. ↩

9.  [https://www.michelleagnes.net/](https://www.michelleagnes.net/), accessed 15th April 2022. ↩

10. [https://vertigo.starts.eu/media/uploads/vertigo-constellactions-residency-public_report.pdf](https://vertigo.starts.eu/media/uploads/vertigo-constellactions-residency-public_report.pdf), accessed 15th April 2022. ↩

11. [http://philippespiesser.com/en](http://philippespiesser.com/en),accessed 15th April 2022. ↩

12. [http://philippespiesser.com/en/projet/skinact-projet-de-recherche/](http://philippespiesser.com/en/projet/skinact-projet-de-recherche/), accessed 15th April 2022. ↩

## Citations

1. Leman, M. (2007). *Embodied music cognition and mediation technology*. MIT press. ↩
2. Godøy, R. I., & Leman, M. (2010). *Musical gestures: Sound, movement, and meaning*. Routledge. ↩
3. Miranda, E. R., & Wanderley, M. M. (2006). *New digital musical instruments: control and interaction beyond the keyboard* (Vol. 21). AR Editions, Inc. ↩
4. Françoise, J., Candau, Y., Fdili Alaoui, S., & Schiphorst, T. (2017). Designing for Kinesthetic Awareness: Revealing User Experiences through Second-Person Inquiry. In *Proceedings of the 2017 CHI Conference on Human Factors in Computing Systems* (pp. 5171–5183). New York, NY, USA: Association for Computing Machinery. Retrieved from [https://doi.org/10.1145/3025453.3025714](https://doi.org/10.1145/3025453.3025714) ↩
5. Jensenius, A. R., & Lyons, M. J. (2017). *A nime reader: Fifteen years of new interfaces for musical expression* (Vol. 3). Springer. ↩
6. Jordà, S., Geiger, G., Alonso, M., & Kaltenbrunner, M. (2007). The ReacTable: Exploring the Synergy between Live Music Performance and Tabletop Tangible Interfaces. In *Proceedings of the 1st International Conference on Tangible and Embedded Interaction* (pp. 139–146). Baton Rouge, Louisiana: Association for Computing Machinery. [https://doi.org/10.1145/1226969.1226998](https://doi.org/10.1145/1226969.1226998) ↩

7. Rasamimanana, N., Bevilacqua, F., Schnell, N., Guedy, F., Flety, E., Maestracci, C., … Petrevski, U. (2010). Modular Musical Objects towards Embodied Control of Digital Music. In *Proceedings of the Fifth International Conference on Tangible, Embedded, and Embodied Interaction* (pp. 9–12). Funchal, Portugal: Association for Computing Machinery. https://doi.org/10.1145/1935701.1935704 ↵

8. West, T., Caramiaux, B., Huot, S., & Wanderley, M. M. (2021). Making Mappings: Design Criteria for Live Performance. In *NIME 2021*. https://doi.org/10.21428/92fbeb44.04f0fc35 ↵

9. Robson, N., Bryan-Kinns, N., & McPherson, A. (2021). On mediating space, sound and experience: interviews with situated sound art practitioners. *Organised Sound: An International Journal of Music and Technology*. ↵

10. Han, J., & Gold, N. (2014). Lessons Learned in Exploring the Leap Motion TM Sensor for Gesture-based Instrument Design. In *NIME*. ↵

11. Yoo, M.-J., Beak, J.-W., & Lee, I.-K. (2011). Creating Musical Expression using Kinect. In *NIME* (pp. 324–325). ↵

12. Sentürk, S., Lee, S. W., Sastry, A., Daruwalla, A., & Weinberg, G. (2012). Crossole: A Gestural Interface for Composition, Improvisation and Performance using Kinect. In *NIME*. ↵

13. Bernardo, F., Arner, N., & Batchelor, P. (2017). O soli mio: exploring millimeter wave radar for musical interaction. In *NIME* (Vol. 17, pp. 283–286). ↵

14. Aylward, R., & Paradiso, J. A. (2006). Sensemble: a wireless, compact, multi-user sensor system for interactive dance. In *Proceedings of the 2006 conference on New interfaces for musical expression* (pp. 134–139). ↵

15. Brown, D., Nash, C., & Mitchell, T. (2018). Simple mappings, expressive movement: a qualitative investigation into the end-user mapping design of experienced mid-air musicians. *Digital Creativity*, *29*(2–3), 129–148. ↵

16. Françoise, J., & Bevilacqua, F. (2018). Motion-sound mapping through interaction: An approach to user-centered design of auditory feedback using machine learning. *ACM Transactions on Interactive Intelligent Systems (TiiS)*, *8*(2), 1–30. ↵

17. Madgwick, S., & Mitchell, T. (2013). x-OSC: A versatile wireless I/O device for creative/music applications. In *SMC Sound and Music Computing Conference*. ↵

18. Medeiros, C. B., & Wanderley, M. M. (2014). A comprehensive review of sensors and instrumentation methods in devices for musical expression. *Sensors*, *14*(8), 13556–13591. ↩

19. Mitchell, T. J., Madgwick, S., & Heap, I. (2012). Musical interaction with hand posture and orientation: A toolbox of gestural control mechanisms. In *Proceedings of the International Conference on New Interfaces for Musical Expression*. ↩

20. Kimura, M., Rasamimanana, N., Bevilacqua, F., Zamborlin, B., Schnell, N., & Fléty, E. (2012). Extracting human expression for interactive composition with the augmented violin. In *International Conference on New Interfaces for Musical Expression (NIME 2012)* (pp. 1–1). ↩

21. Fernandez, J. M., Köppel, T., Verstraete, N., Lorieux, G., Vert, A., & Spiesser, P. (2017). Gekipe, a gesture-based interface for audiovisual performance. In *NIME* (pp. 450–455). ↩

22. Bermudez, B., Delahunta, S., Marijke, H., Chris, Z., Bevilacqua, F., Alaoui, S. F., & Gutierrez, B. M. (2011). The double skin/double mind interactive installation. *The Journal for Artistic Research*. ↩

23. Candau, Y., Françoise, J., Alaoui, S. F., & Schiphorst, T. (2017). Cultivating kinaesthetic awareness through interaction: Perspectives from somatic practices and embodied cognition. In *Proceedings of the 4th International Conference on Movement Computing* (pp. 1–8). ↩

24. Bevilacqua, F., Guédy, F., Schnell, N., Fléty, E., & Leroy, N. (2007). Wireless Sensor Interface and Gesture-Follower for Music Pedagogy. In *Proceedings of the 7th International Conference on New Interfaces for Musical Expression* (pp. 124–129). New York, New York: Association for Computing Machinery. https://doi.org/10.1145/1279740.1279762 ↩

25. Zappi, V., & McPherson, A. P. (2014). Dimensionality and Appropriation in Digital Musical Instrument Design. In *NIME* (Vol. 14, pp. 455–460). ↩

26. van Vugt, F. T., & Ostry, D. J. (2019). Early stages of sensorimotor map acquisition: learning with free exploration, without active movement or global structure. *Journal of Neurophysiology*, *122*(4), 1708–1720. ↩

27. Braun, V., & Clarke, V. (2006). Using thematic analysis in psychology. *Qualitative Research in Psychology*, *3*(2), 77–101. ↩

28. Tanaka, A., Parkinson, A., Settel, Z., & Tahiroglu, K. (2012). A Survey and Thematic Analysis Approach as Input to the Design of Mobile Music GUIs. In *Proceedings of the International Conference on New Interfaces for Musical Expression*. Ann Arbor, Michigan: University of Michigan. https://doi.org/10.5281/zenodo.1178431 ↩

29. Braun, V., & Clarke, V. (2021). Can I use TA? Should I use TA? Should I not use TA? Comparing reflexive thematic analysis and other pattern-based qualitative analytic approaches. *Counselling and Psychotherapy Research*, *21*(1), 37–47. ↩

30. Dobrian, C., & Koppelman, D. (2006). The 'E' in NIME: Musical Expression with New Computer Interfaces. In *NIME* (Vol. 6, pp. 277–282). ↩

31. Gurevich, M., & Trevi~no, Jeffrey. (2007). Expression and its discontents: toward an ecology of musical creation. In *Proceedings of the 7th international conference on New interfaces for musical expression* (pp. 106–111). ↩

32. Gurevich, M., & Fyans, A. C. (2011). Digital musical interactions: Performer–system relationships and their perception by spectators. *Organised Sound*, *16*(2), 166–175. ↩

33. Reeves, S., Benford, S., O'Malley, C., & Fraser, M. (2005). Designing the spectator experience. In *Proceedings of the SIGCHI conference on Human factors in computing systems* (pp. 741–750). ↩

34. Capra, O., Berthaut, F., & Grisoni, L. (2020). Have a SEAT on Stage: Restoring Trust with Spectator Experience Augmentation Techniques. In *Proceedings of the 2020 ACM Designing Interactive Systems Conference* (pp. 695–707). New York, NY, USA: Association for Computing Machinery. Retrieved from https://doi.org/10.1145/3357236.3395492 ↩

35. Calegario, F., Tragtenberg, J., Frisson, C., Meneses, E., Malloch, J., Cusson, V., & Wanderley, M. M. (2021). Documentation and Replicability in the NIME Community. In *NIME 2021*. https://doi.org/10.21428/92fbeb44.dc50e34d ↩

36. Zayas-Garin, L., Harrison, J., Jack, R., & McPherson, A. (2021). DMI Apprenticeship: Sharing and Replicating Musical Artefacts. In *NIME 2021*. https://doi.org/10.21428/92fbeb44.87f1d63e ↩

37. Pigrem, J., & McPherson, A. P. (2018). Do We Speak Sensor? Cultural Constraints of Embodied Interaction. In *NIME* (pp. 382–385). ↩

38. Hayes, L. (2019). Beyond skill acquisition: Improvisation, interdisciplinarity, and enactive music cognition. *Contemporary Music Review*, *38*(5), 446–462. ↩

39. Altavilla, A., Caramiaux, B., & Tanaka, A. (2013). Towards Gestural Sonic Affordances. In *NIME* (pp. 61–64). ↩

40. Pascual, B. B. (2016). Double Skin/Double Mind: Emio Greco| PC's interactive installation. In *Transmission in Motion* (pp. 115–122). Routledge. ↩

41. Matuszewski, B., Larralde, J., & Bevilacqua, F. (2018). Designing movement driven audio applications using a web-based interactive machine learning toolkit. In *Web Audio Conference (WAC).* ↩

42. Rasamimanana, N., Bevilacqua, F., Schnell, N., Guedy, F., Flety, E., Maestracci, C., … Petrevski, U. (2010). Modular musical objects towards embodied control of digital music. In *Proceedings of the fifth international conference on Tangible, embedded, and embodied interaction* (pp. 9–12). ↩